\documentclass[aps,prl,twocolumn,superscriptaddress,tightenlines,floatfix]{revtex4}
\usepackage{epsfig,rotating}

\newcommand{\nuc}[2]{\hbox{$^{#1}$#2}}
\begin{document}
\bibliographystyle{apsrev}


\title{Cross-shell excitation in two-proton knockout: Structure of
$^{52}$Ca }
\author{A.\ Gade}
    \affiliation{National Superconducting Cyclotron Laboratory,
      Michigan State University, East Lansing, Michigan 48824}
     \affiliation{Department of Physics and Astronomy,
      Michigan State University, East Lansing, Michigan 48824}
\author{R.V.F.\ Janssens}
    \affiliation{Physics Division, Argonne National Laboratory, Argonne,
      IL 60439}
\author{D.\ Bazin}
    \affiliation{National Superconducting Cyclotron Laboratory,
      Michigan State University, East Lansing, Michigan 48824}
\author{R.\ Broda}
    \affiliation{Institute of Nuclear Physics, Polish Academy of
      Science, PL-31342 Cracow, Poland}
\author{B.\,A.\ Brown}
    \affiliation{National Superconducting Cyclotron Laboratory,
      Michigan State University, East Lansing, Michigan 48824}
    \affiliation{Department of Physics and Astronomy,
      Michigan State University, East Lansing, Michigan 48824}
\author{C.\,M.~Campbell}
    \affiliation{National Superconducting Cyclotron Laboratory,
      Michigan State University,
      East Lansing, Michigan 48824}
    \affiliation{Department of Physics and Astronomy,
      Michigan State University, East Lansing, Michigan 48824}
\author{M.\,P.\ Carpenter}
    \affiliation{Physics Division, Argonne National Laboratory, Argonne,
      IL 60439}
\author{J.\,M.\ Cook}
    \affiliation{National Superconducting Cyclotron Laboratory,
      Michigan State University, East Lansing, Michigan 48824}
    \affiliation{Department of Physics and Astronomy,
      Michigan State University, East Lansing, Michigan 48824}
\author{A.\,N. Deacon}
    \affiliation{School of Physics and Astronomy, Schuster Laboratory,
      University of Manchester, Manchester M13 9PL, United Kingdom}
\author{D.-C.\ Dinca}
    \affiliation{National Superconducting Cyclotron Laboratory,
      Michigan State University, East Lansing, Michigan 48824}
    \affiliation{Department of Physics and Astronomy,
      Michigan State University, East Lansing, Michigan 48824}
\author{B.\ Fornal}
     \affiliation{Institute of Nuclear Physics, Polish Academy of
      Science, PL-31342 Cracow, Poland}
\author{S.\,J.\ Freeman}
    \affiliation{School of Physics and Astronomy, Schuster Laboratory,
      University of Manchester, Manchester M13 9PL, United Kingdom}
\author{T.\ Glasmacher}
    \affiliation{National Superconducting Cyclotron Laboratory,
      Michigan State University, East Lansing, Michigan 48824}
    \affiliation{Department of Physics and Astronomy,
      Michigan State University, East Lansing, Michigan 48824}
\author{P.\,G.\ Hansen}
    \affiliation{National Superconducting Cyclotron Laboratory,
      Michigan State University,
      East Lansing, Michigan 48824}
    \affiliation{Department of Physics and Astronomy,
      Michigan State University, East Lansing, Michigan 48824}
\author{B.\,P.\ Kay}
    \affiliation{School of Physics and Astronomy, Schuster Laboratory,
      University of Manchester, Manchester M13 9PL, United Kingdom}
\author{P.\,F.\ Mantica}
    \affiliation{National Superconducting Cyclotron Laboratory,
      Michigan State University,
      East Lansing, Michigan 48824}
    \affiliation{Department of Chemistry, Michigan State University,
      East Lansing, MI 48824}
\author{W.\,F.\ Mueller}
    \affiliation{National Superconducting Cyclotron Laboratory,
      Michigan State University, East Lansing, Michigan 48824}
\author{J.\,R.\ Terry}
    \affiliation{National Superconducting Cyclotron Laboratory,
      Michigan State University,
      East Lansing, Michigan 48824}
    \affiliation{Department of Physics and Astronomy,
      Michigan State University, East Lansing, Michigan 48824}
\author{J.\,A.\ Tostevin}
    \affiliation{Department of Physics, School of Electronics and
      Physical Sciences, University of Surrey, Guildford, Surrey GU2 7XH,
      United Kingdom}
\author{S.\ Zhu}
    \affiliation{Physics Division, Argonne National Laboratory, Argonne,
      IL 60439}
\date{\today}

\begin{abstract}
The two-proton knockout reaction $^9$Be($^{54}$Ti,$^{52}$Ca$ +
\gamma$) has been studied at 72~MeV/nucleon. Besides the
strong feeding of the $^{52}$Ca ground state, the only other
sizeable cross section proceeds to a 3$^-$ level at 3.9 MeV.
There is no measurable direct yield to the first excited 2$^+$ state
at 2.6 MeV. The results illustrate the potential of such direct
reactions for exploring cross-shell proton excitations in neutron-rich
nuclei and confirms the doubly-magic nature of $^{52}$Ca.
\end{abstract}

\maketitle

For decades, the cornerstone of nuclear structure has been the concept 
of single-particle motion in a well-defined potential leading to 
shell structure and magic numbers governed by the strength
of the mean-field spin-orbit interaction~\cite{May49}.
Recent observations in exotic, neutron-rich nuclei 
have demonstrated that the sequence and
energy spacing of single-particle orbits is not as immutable
as once thought: some of the familiar magic numbers
disappear and new shell gaps develop~\cite{Bro01}. Cross-shell excitations, 
arising from the promotion of nucleons across shell gaps,
probe changes in shell structure. They are, however,
not always readily identifiable in nuclear spectra. 
This letter demonstrates that two-proton knockout
reactions can examine, selectively, cross-shell {\it
proton} excitations in {\it neutron}-rich systems.

Single-nucleon knockout reactions with fast radioactive beams are established
tools to investigate the properties of halo nuclei \cite{Ha:95} and to
study beyond 
mean-field correlations, indicated by the quenching of
spectroscopic strengths \cite{Ga:04}. Eikonal theory \cite{Ha:03}
provides a suitable framework for the extraction of quantitative
nuclear structure information from such reactions. In contrast, the
potential of 
two-nucleon knockout as a spectroscopic tool has been recognized
only recently. Bazin {\it et al.} \cite{Ba:03} have shown
that two-proton removal reactions from beams of neutron-rich
species at intermediate energies proceed as direct reactions and
that partial cross sections to different
final states of the residue provide structure information. More
recently, such a reaction was used to infer the magicity of the
very neutron-rich $^{42}$Si nucleus \cite{Fr:05}.

In the current experiment, sizable cross sections for the
$^9$Be($^{54}$Ti,$^{52}$Ca$ + 
\gamma)X$ reaction were found to feed only the $^{52}$Ca ground
state and a 3$^-$ level with an excitation energy near 4 MeV,
bypassing completely the first 2$^+$ level at 2.6 MeV. These
observations can be reproduced qualitatively by calculations 
which assign the 3$^-$ level to the promotion of protons across 
the $Z=20$ shell gap. In addition, the data confirm the 
presence of a neutron sub-shell closure at $N=32$, the subject
of much recent attention \cite{Ots:01,Pri:01,Huc:85,Jan:02,Lid:04,Di:05}.

The $^{54}$Ti secondary ions were produced by fragmentation of a
130~MeV/nucleon $^{76}$Ge beam, delivered by the Coupled
Cyclotron Facility of the National Superconducting Cyclotron
Laboratory, onto a $^9$Be fragmentation
target. The ions were selected in the A1900 large-acceptance
fragment separator \cite{a1900}, which was operated with two
settings during different phases of the experiment; 1\% momentum
acceptance and no momentum restriction, respectively. The secondary
beam, with a mid-target energy of 72  
MeV/nucleon, interacted with another, 375(4)~mg/cm$^2$ thick,
$^{9}$Be foil located at the target position of the
high-resolution S800 spectrograph \cite{s800}. The knockout
residues were identified event-by-event from the time of flight
between scintillators monitoring the beam, the energy loss
measured in an ionization chamber, and the position and angle
information provided by the cathode-readout drift counters in the
S800 focal plane~\cite{s800}. The $^9$Be reaction target was
surrounded by SeGA, an array of seventeen 32-fold segmented HPGe
detectors~\cite{sega}, arranged in two rings (90$^\circ$ and
37$^\circ$ with respect to the beam axis).  

An inclusive cross section for two-proton knockout to all final
states was derived from the yield of detected $^{52}$Ca residues
and the number of incoming $^{54}$Ti projectiles, taking into
account the density of the secondary $^9$Be target. To minimize
effects arising from acceptance limitations in the S800
spectrograph, only data with the momentum acceptance of the A1900
fragment separator reduced to 1\% were used to deduce the cross
section. Systematic uncertainties accounting for particle identification (6\%),
stability and purity of the beam (5\%), and the momentum acceptance of the
spectrograph (4\%) were added to the statistical uncertainty in
quadrature.

Figure 1 presents the inclusive parallel-momentum distribution for
all two-proton knockout events from $^{54}$Ti to $^{52}$Ca,
together with that of the unreacted $^{54}$Ti projectiles passing
through the target. As discussed below, $\sim$ 70\% of the two-proton
knockout cross section 
feeds the $^{52}$Ca ground state directly, making
it worthwhile to consider these inclusive data first.
The $^{52}$Ca momentum distribution of
Fig. 1 is fairly narrow and centered close to the beam velocity.
As argued in Ref.~\cite{Ba:03}, this observation is consistent
with the reaction mechanism being direct in nature. The magnitude
of the inclusive cross section, $\sigma_{inc}=0.32(4)$~mb, is of
the same order as that reported in Ref.~\cite{Fr:05} for
two-proton knockout to $^{42}$Si (0.12(3) mb). It is also
significantly smaller than that observed for the $^9$Be($^{28}
$Mg,$^{26}$Ne) reaction~\cite{Ba:03}, for example. To first order,
this indicates that the number of valence protons available for
the process must be small \cite{Ba:03}.

\begin{figure}[h]
\epsfxsize 7.5cm \epsfbox{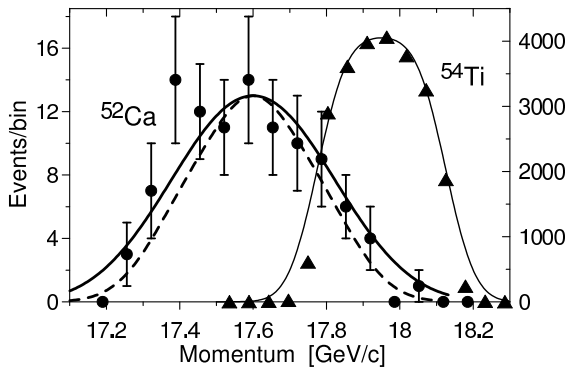} \caption{\label{fig:momentum}
Longitudinal momentum distribution of the $^{52}$Ca knockout
residues and the unreacted $^{54}$Ti
projectiles (momentum acceptance of the A1900 restricted to 1\%). The
$^{52}$Ca data are compared to a 
calculated shape given by the convolution of the separate
distributions for two uncorrelated protons removed from the
$f_{7/2}$ orbit (dashed line) and to the same curve, but folded
with the momentum profile of the $^{54}$Ti beam (solid line).}
\end{figure}

To proceed further in understanding the momentum distribution, a
description of $^{54}$Ti as a semi-magic nucleus, with a closed
$N=32$ neutron shell and two $f_{7/2}$ valence protons outside the
$Z=20$ core, has been adopted, in keeping with recent experimental
information on the level structure and the $B(E2;0^+\rightarrow
2^+_1)$ transition rate of this nucleus~\cite{Jan:02,Di:05}.
Following Ref.~\cite{Ba:03}, the dashed curve for $^{52}$Ca in
Fig.~1 is the result of calculations where the momentum
distributions for the removal of two independent, i.e.
uncorrelated, $f_{7/2}$ protons have been convoluted. The solid
line results when this distribution is folded with the
momentum profile of the unreacted $^{54}$Ti projectiles, to
account for the momentum spread in the incoming beam and the
straggling in the target. The agreement is satisfactory, and the
measured momentum distribution consistent with the removal of two
$f_{7/2}$ protons being the dominant reaction process.

The $\gamma$-ray spectrum measured in
coincidence with the $^{52}$Ca residues (Fig. 2) displays two
transitions with energies of 1430(7) and 2562(13) keV. The latter can
be identified with the 2563(1) 
keV line reported by Huck {\it et al}~\cite{Huc:85}, the only
prominent line seen in the $\beta$ decay of $^{52}$K to $^{52}$Ca,
and assigned by these authors as the 2$^+$$\rightarrow$0$^+$
ground-state transition. The observation was confirmed in a more
recent $\beta$-decay study~\cite{PhD}. At present, this second
study is also the only one to report additional transitions in
$^{52}$Ca. In particular, a 3990 keV state was found, with the
second highest feeding in the $\beta$-decay process, and with
deexcitation towards the $2^+$ level via a 1427(1) keV transition.
It is tempting to associate this $\gamma$ ray with the
1430(7) keV line of Fig.~2. In the work of Ref.\ \cite{PhD}, the
3990 keV state was assigned $J^{\pi}=3^-$ as the most likely quantum
numbers based on the measured feeding in the $\beta$-decay
process, the feeding from another level at 5951 keV, and the
deexcitation pattern towards the $2^+$ state. In addition, the
excitation energy of the 
state at 3990 keV is close to that of the known first $3^-$
excitation along the even-even Ca isotopic chain illustrated in
Fig.~3.

\begin{figure}[h]
\epsfxsize 7.5cm \epsfbox{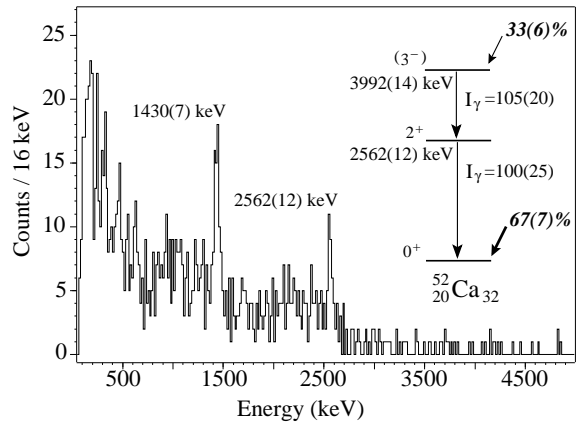} \caption{\label{fig:level}
$\gamma$-ray spectrum (2.75\% detection efficiency at 1~MeV) in
  coincidence with the $^{52}$Ca knockout residues.  
The two $\gamma$-ray transitions establish the cascade shown in
the inset. The arrows indicate the feeding profile by the reaction
($\sigma(3^-)=0.11(3)$~mb and $\sigma(0^+)=0.21(3)$~mb). For this part
of the experiment, the A1900 fragment separator was operated without
momentum restriction.} 
\end{figure}

\begin{figure}[h]
\epsfxsize 8.0cm \epsfbox{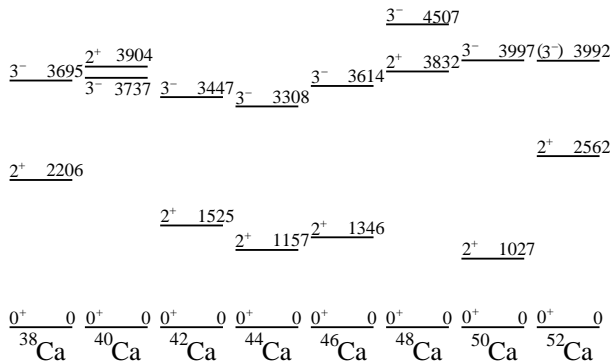} \caption{\label{fig:spectrum}
Systematics of the first $2^+$ and $3^-$ excitations along the
even-even Ca isotopic chain (data from Ref.~\cite{PhD,ISO,Bro:05}).}
\end{figure}

There is no
appreciable direct feeding of the $2^+$ level in the two-proton
knockout process as the 1430(7) and 2562(13) keV lines are
measured to have the same relative intensity 
(see Fig. 2). As is discussed below, this
$2^+$ state is understood as a neutron particle-hole excitation
across the $N=32$ shell gap, involving the $p_{3/2}$ and $p_{1/2}$
orbitals. A priori, such an excitation is not expected to be
populated appreciably in the reaction of a $^{54}$Ti projectile
with two $f_{7/2}$ valence protons and no valence neutrons. The
same reasoning leads to the complementary conclusion
that the sizable cross section to the 3392(14) keV state is a
direct indication of a proton excitation of 3992(14) keV. Taking
account of the closed-shell character of the $^{54}$Ti projectile 
and of the $^{52}$Ca residue, this suggests that the two-proton
knockout mechanism will be an important tool for exploring
cross-shell proton excitations in neutron-rich systems.

Clearly, this first principles reasoning requires substantiation
by more detailed calculations. While the sudden, direct nature of
the two-proton removal reaction was identified in Ref.
\cite{Ba:03}, the associated two-proton structures and
reaction mechanism were treated only approximately. A more
complete formalism, using eikonal theory and shell-model wave 
functions, has been developed since for the two-nucleon stripping 
(inelastic breakup) mechanism \cite{tos:04}.
This formalism has now been extended 
to include a complete calculation of removal events in which one proton is
absorbed (stripped) and a second is elastically dissociated
(diffracted) by the target. The (smaller) cross section for 
elastic dissociation of both (strongly-bound) protons is 
estimated as well. Details will be presented elsewhere
\cite{tos:06}. Of relevance here is that, for four test-case
$sd$-shell  
nuclei, the calculated inclusive cross sections overpredict the 
measured values by a factor of two \cite{tos:06}, requiring a suppression 
factor, $R_s(2N) \approx 0.5$, of the shell-model strengths: the 
analog of the well-documented suppression seen in nuclear 
and electron-induced single-nucleon knockout reactions \cite{dic:04}.

The eikonal $S$-matrices were calculated from the residue and
target one-body matter densities using the optical limit of
Glauber's multiple scattering theory \cite{Gla59,Tos01}. A
Gaussian nucleon-nucleon (NN) effective interaction was assumed
\cite{Tos99} with a range of 0.5~fm and a strength determined, in
the usual way, by the free pp and np cross sections and the
real-to-imaginary ratios of the forward NN scattering amplitudes 
\cite{Ray79}. The $^9$Be density was of Gaussian form with a 
{\it rms} matter radius of 2.36 fm \cite{Oza01}. The density 
of $^{52}$Ca was taken from a spherical Skyrme (SkX) Hartree-Fock 
(HF) calculation \cite{skx}, with a {\it rms} matter radius of 3.632 fm.

The protons were assumed to occupy $fp$- and $sd$-shell
model orbitals with spectroscopic amplitudes obtained using 
the code {\sc OXBASH} \cite{oxbash}. Their single-particle 
wave functions were calculated in Woods-Saxon potential wells with 
fixed diffuseness ($a=0.70$ fm) and radius parameters 
($r_0$) adjusted to reproduce the {\it rms} radii of the
proton single-particle states given by the HF calculation above~\cite{Ga:04}. 
The strengths of the binding potentials were adjusted to support 
bound eigenstates with the physical ground- and excited-state 
separation energies. The ground-state to ground-state two-proton
separation energy for $^{54}$Ti was calculated to be $S_{2p}=27.83$~MeV,
in agreement with the measured $S_{2p}=27.66(71)$~MeV~\cite{Aud03}.

The nuclear structure input to the cross section calculations was 
also based on the following additional considerations. While neutron
excitations of $fp$-shell nuclei have been the subject of much 
recent attention and modern effective interactions have been developed 
that reproduce the data satisfactorily 
\cite{Ots:01,Pri:01, Jan:02,Lid:04,Di:05}, the same cannot yet be 
said for the proton excitations relevant for the present work. 
Here, calculations were carried out with the modified WBMB Hamiltonian 
of Ref.\cite{War91}, which is able to describe the low-level structure
of \nuc{50}{K}  
and the rather unusual characteristics of its $\beta$ decay to \nuc{50}Ca. 
This interaction results in a $0^-$ \nuc{50}K ground state 
and a $2^-$ excitation at 550 keV. The situation is reversed 
in \nuc{52}K, and the $2^-$ level becomes the ground state 
while the $0^-$ excitation is calculated to be located at 2.3 MeV. 
These \nuc{52}K ground state $I^{\pi}$ quantum numbers agree with the values
inferred in the most recent $\beta$-decay study~\cite{PhD}. The
calculations also account for the observed feeding of $^{52}$Ca  
levels through (i) a first forbidden transition to the $2_1^+$ level
and (ii) a Gamow Teller
(GT) transition to the $3^-$ state, as well as for the absence of 
direct feeding of the ground state (first forbidden GT transition). 

Besides the \nuc{52}Ca ground state, the cross section calculations
also required proton wave functions for the $3^-$ level and for
positive-parity 
excitations associated with two holes in the $sd$ shell. The WBMB interaction
places the negative-parity states $\sim$~3 MeV above the experimental value, 
illustrating the difficulties with the interaction alluded to above. 
(Attempts with the interaction 
of Ref.~\cite{Num01} result in a similar problem). Nevertheless, the
negative-parity \nuc{52}Ca spectrum has a $3^-$ state as its lowest
excitation. The need to consider two particle-two hole 
excitations in two-proton knockout comes from the observation that, 
in \nuc{48}Ca, a 4.28 MeV, $0^+$ state with this intrinsic structure
(see Fig. 1 in \cite{skx}) lies within 225 keV of the $3^-$ level. The 
presence of a similar level in the same energy range in \nuc{52}Ca
cannot be ruled out. Unfortunately, its 
excitation energy cannot be readily estimated~\cite{footnote}, 
and in this case as well the cross-shell
calculations can only be viewed as qualitative.           

With the wave functions derived in this manner, the calculated
$fp$-shell, two-proton knockout partial cross section  
to the \nuc{52}Ca $0^+$ ground state is 0.38 mb. Comparison with
the measured value, 0.21(3) mb, yields the ratio $R_s(2N)=0.55(8)$, in 
agreement with the available $sd$-shell systematics \cite{tos:06}.
The direct population of the $2_1^+$ state is predicted to
be very small as this level corresponds to a neutron excitation: the 
calculated value of 0.04 mb is well below the detection sensitivity of the
present measurements.

In view of the uncertainties discussed above,
the knockout cross section calculations to the other excited states
were restricted to the simplest shell-model configurations expected. Thus, 
$3^-$ states with (i) one-proton hole in the $d_{3/2}$ shell and 
(ii) one-proton hole in the $s_{1/2}$ state were considered (i.e., 
the configurations were of the type $\pi((d_{3/2})^{-1} \times
(f_{7/2})^{+1})$, 
for example). Similarly, for the $0^+$ states, configurations 
with (iii) two $d_{3/2}$  proton holes and (iv) two $s_{1/2}$ proton holes
were included. The calculated two-proton knockout cross sections
(taking $R_s(2N)=0.55$)  
are 0.13, 0.19, 0.09 and 0.09~mb, respectively. As the physical lowest 
$3^-$ state is expected to correspond to a linear combination of
configurations (i) and (ii), their sum, 0.32~mb, can be used as a 
upper limit of the two-proton knockout cross
section, assuming coherent superposition. 
A similar upper limit to the $0^+_2$ level is then 0.18~mb. 

An interpretation of the experimental results is, therefore, 
that the 3990~keV level can be associated with the $3^-$
state seen in \nuc{52}{K} $\beta$ decay, and that the two-proton
knockout cross section of 0.11(3) mb is several times weaker than 
the maximum allowed by the $3^-$ calculation. The absence of any other
notable strength in the experimental spectrum then suggests that the 
$0^+_2$ state lies above the neutron threshold of 4.7(7)~MeV.
In addition, it is worth noting that the two-proton knockout 
cross sections summed over all one- and two-proton hole states 
in the $sd$-shell are 1.93 and 9.51~mb, respectively. Most of
these strengths proceed towards levels above the 
neutron separation energy and lead to states in 
lighter Ca isotopes that cannot be discerned by the experiment.

In summary, the present work has demonstrated that two-proton knockout
reactions represent an important tool in the arsenal of experimental
techniques aimed at the study of neutron-rich nuclei. The specific 
$^9$Be($^{54}$Ti,$^{52}$Ca$ +\gamma)X$ reaction investigated here 
was instrumental in identifying a proton cross-shell 
excitation in an exotic nucleus with large neutron excess. In addition,
the data provided confirmation of the $N=32$ sub-shell closure in 
$^{54}$Ti. Finally, the reduction of the spectroscopic strength,
verified thus far only for $sd$-shell nuclei ($R_s(2N) \approx 0.5$)
was found to apply to this $fp$-shell nucleus as well.

\begin{acknowledgments}
This work was supported by the National Science Foundation under
Grants No. PHY-0110253, PHY-9875122, and PHY-0244453, by the
U.S. Department of Energy, Office of Nuclear Physics, under 
Contract No. W31-109-ENG-38, the UK Engineering and Physical
Sciences Research Council (grant EP/D003628) and
by Polish Scientific Committee Grant No. 1P03B 059 29.
\end{acknowledgments}


\begin{thebibliography}{10}
\bibitem{footnote} An estimate of this $0^+_2$
excitation energy can be obtained from
binding energies (BE) through   
[2$BE$(\nuc{52}{Ca})-$BE$(\nuc{54}{Ti})-$BE$(\nuc{50}{Ar})]+4$\bar{V}$ 
with the residual proton-neutron interaction $\bar{V}=-0.75$~MeV derived from  
this expression for the $0^+_2$ level in \nuc{48}Ca. Unfortunately,
the BE values of the neutron-rich nuclei involved are
not known with the required accuracy.
\end{thebibliography}
\end{document}